\documentclass{PoS}

\title{Recent Progress in Lattice QCD Thermodynamics }

\ShortTitle{QCD Thermodynamics }

\author{\speaker{Carleton DeTar} \\
        Physics Department, University of Utah, Salt Lake City, UT 84112, USA\\
        E-mail: \email{detar@physics.utah.edu}}

\abstract{ This review gives a critical assessment of the current
  state of lattice simulations of QCD thermodynamics and what it
  teaches us about hot hadronic matter.  It outlines briefly lattice
  methods for studying QCD at nonzero temperature and zero baryon
  number density with particular emphasis on assessing and reducing
  cutoff effects.  It discusses a variety of difficulties with methods
  for determining the transition temperature.  It uses results reported
  recently in the literature and at this conference for
  illustration, especially those from a major study carried out by
  the HotQCD collaboration.}

\FullConference{The XXVI International Symposium on Lattice Field Theory \\
		 July 14 - 19, 2008\\
		 Williamsburg, Virginia, USA}

\newcommand{\Tr}{{\rm Tr}}

\newcommand{\bi}{\begin{itemize}}
\newcommand{\ei}{\end{itemize}}
\newcommand{\be}{\begin{equation}}
\newcommand{\ee}{\end{equation}}
\newcommand{\bea}{\begin{eqnarray}}
\newcommand{\eea}{\end{eqnarray}}
\newcommand{\ie}{{\it i.e.}}
\newcommand{\eg}{{\it e.g.}}
\newcommand{\vs}{{\it vs.\ }}

\begin{document}

\section{Introduction}

At an early stage the universe was very likely a quark-gluon plasma.
In heavy ion colliders we seek to recreate this state of matter and
study its properties.  Lattice gauge theory is ideally suited for the
fully nonperturbative study of quantum chromodynamics under conditions
close to thermal equilibrium.  The insights gained from lattice
simulations can be extrapolated through hydrodynamic modeling to the
quasi-equilibrium expansion of the plasma.  In this way lattice
calculations provide crucial assistance in the interpretation of
experimental results \cite{Hwa:2004yg}.

The baryon density was essentially zero in the early universe.  At
high densities, even at low temperature, popular tradition predicts
that hadronic matter is also in a deconfined plasma state.  At still
higher densities even more unusual phases, such as a color-flavor
locked phase have been proposed \cite{Alford:1998mk}.  For technical
reasons, such high densities are beyond the reach of lattice
simulations using standard methods.  At least we can hope to simulate
matter at the low densities found in heavy ion collisions.

In this talk I consider only progress in calculations at zero baryon
number density.  In the companion talk Shinji Ejiri describes
developments in nonzero density calculations \cite{Ejiri:Lat2008}.

Here are highlights of recent advances at zero baryon density, which
I will cover in this talk:
\bi
\item HotQCD study.  A high statistics study is being carried out on
  the IBM BlueGene at the Lawrence Livermore National Laboratory.
  This study compares results in closely matched simulations from two
  staggered fermion actions, namely asqtad and p4fat3, and it is
  providing the first large-scale simulation at $N_\tau = 8$ with
  domain wall fermions.  (See talks by M.~Cheng \cite{Cheng:Lat2008},
  R.~Gupta \cite{Gupta:Lat2008}, and W.~Soeldner
  \cite{Soeldner:Lat2008}.)
\item Chiral susceptibility.  New insights into the behavior of the
  chiral susceptibility will change the determination of $T_c$ using
  this quantity \cite{Karsch:2008ch}.
\item Equation of state.  A new method has been proposed \cite{Umeda:Lat2008}.
\item Transport coefficients.  There are new ideas and methods for
  computing them \cite{Meyer:Lat2008}.
\item Spatial string tension.  A new result agrees surprisingly well with
  3D perturbation theory \cite{Cheng:2008bs}.
\ei
I will not have time to cover interesting studies of QCD-like theories
with a large number of flavors \cite{Deuzeman:Lat2008}.  And I regret
that time and space did not permit covering all recent work in this field.

I will try to give a general and fairly critical overview, using
selected results from the parallel sessions as illustrations and
leaving the details to the parallel sessions.  After a brief review of
lattice methodology I discuss potential cutoff problems with various
actions and focus on issues and confusion in determining $T_c$.
Turning to results, I highlight some new methods and results for the
equation of state, allude to recent progress in determining transport
coefficients, and end with mention of a little surprise concerning
predictions of dimensional reduction for the spatial string tension.

\section{Lattice Methodology}

Lattice methods are especially well suited for simulating a quantum
statistical ensemble in thermal equilibrium at fixed temperature $T$.
We set a finite imaginary time interval 
\be
   aN_\tau = 1/T
\ee
for lattice spacing $a$ and $N_\tau$ sites in imaginary time, and we
impose periodic (antiperiodic) boundary conditions on the bosonic
(fermionic) fields.  Under these conditions the lattice Feynman path
integral generates the quantum partition function for the underlying
hamiltonian $H$,
\be
   Z = \Tr \exp(-H/T),
\label{eq:partfunc}
\ee
in the continuum limit.  Operator expectation values are thermal
expectation values in this ensemble.  Since simulations with standard
methods are limited to equilibrium and near-equilibrium processes, to
apply lattice results to the nonequilibrium conditions of heavy ion
collisions requires phenomenological modeling.

The temperature is varied by changing either $N_\tau$ or $a$.  The
latter strategy is more common.  At fixed $N_\tau$, decreasing the
gauge coupling $g^2$ decreases $a$, so $T$ grows.  It is common now to
adjust the bare lattice quark masses together with the lattice spacing
so that zero temperature meson masses remain fixed.  In this way
variations in observables can be attributed to changes in temperature
and not also to changes in the Hamiltonian. Such trajectories through
parameter space are called ``lines of constant physics.''

Of course, to connect with reality we need also to take the continuum
limit.  With the fixed $N_\tau$ strategy, the lattice is coarser at
low temperatures and finer at high temperatures.  For a given
temperature, obviously, we approach the continuum by repeating the
calculation at smaller $a$ and larger $N_\tau$.  Contemporary lattice
simulations have $N_\tau$ as large as 12 for some quantities
\cite{Szabo:Lat2008}, but 6 and 8 are typical for expensive quantities
such as the equation of state.  Table \ref{tab:Ntvsa} shows the
relationship between lattice spacing and $N_\tau$ at $T = 180$ MeV,
near the crossover temperature for QCD.  We see that by standards of
contemporary zero temperature simulations, thermodynamic simulations
at $N_\tau = 6$ and 8 are rather coarse at this temperature, and
$N_\tau = 4$ is extremely coarse.  Thus we must be alert to the
possibility of distortions due to cutoff effects.
\begin{table}[hb]
   \begin{center}
     \begin{tabular}{|l|l|l|l|l|l|}
     \hline
      $N_\tau$ & 4   &  6   & 8   & 10   & 12 \\
     \hline
      $a$ (fm)      & 0.27 &  0.18 & 0.14 & 0.11 & 0.09\\
     \hline
     \end{tabular}
   \end{center}
\caption{Lattice spacing \vs $N_\tau$ at $T = 180$ MeV.}
\label{tab:Ntvsa}
\end{table}

The continuum limit can be expensive.  For the equation of state the
computational cost grows with decreasing lattice spacing as $a^{-11}$.
This places a high premium on reducing undesirable cutoff effects at a
coarse lattice spacing.  The degree of ``improvement'' of the lattice
action is a significant factor.

The most extensive recent simulations use staggered fermions with
varying degrees of improvement.  The asqtad staggered formalism is
designed to eliminate cutoff effects at ${\cal O}(a^2)$, leaving
errors at ${\cal O}(\alpha_s a^2)$ (see references in
\cite{Bernard:2001av}).  The p4fat3 staggered action is also improved,
but it does not eliminate all such effects \cite{Heller:1999xz}.  Both
actions also improve the free-quark dispersion relation, which is
desirable in a high temperature deconfined environment.  The
Budapest-Wuppertal action with stout gauge links and unimproved
staggered fermions does not improve the dispersion relation
\cite{Aoki:2005vt}, but it does reduce taste-splitting effects.

Improvement is good, but one may carry it too far.  Improvement tends
to fatten action operators, in which case localization could become an
issue.  It is plausible that the extent to which a lattice simulation
approximates the quantum partition function (\ref{eq:partfunc})
depends on the locality of the lattice transfer matrix.  Ideally the
localization length of the action $\ell$ should be much less than
$1/T$ or in lattice units, much less than $N_\tau$.

For free fermions cutoff effects for various lattice formulations can
be studied analytically.  Recently Hegde {\it et al.}
\cite{Hegde:2008nx} looked at deviations from the expected
free-fermion Stefan-Boltzmann relation for the pressure $p$ as a
function of $1/N_\tau^2$ (equivalently $a^2$) and chemical potential
$\mu/T$:
\be
       \frac{p}{T^4} = 
           \sum_{k=0}^\infty A_{2k} P_{2k}(\mu/\pi T) 
                  \left(\frac{\pi}{N_\tau}\right)^{2k}
\label{eq:Pexpansion}
\ee
The leading term $A_0$ is the Stefan-Boltzmann term.  The ratios of
higher coefficients $A_{2k}/A_0$ measure the strength of the cutoff
effects.  These terms measure the ability of the action to approximate
the continuum free fermion dispersion relation.  Table \ref{tab:SB}
reproduces their results for a variety of actions.  We see that the
hypercube action \cite{Bietenholz:1996pf} has pleasingly small
coefficients.  The Naik (asqtad) and p4 (p4fat3) actions remove the
second order term as designed, but the p4 action is better at sixth
order.  The standard (unimproved) staggered action (regardless of
gauge-link smearing) does as poorly as the standard (and
clover-improved) Wilson actions.  The overlap and domain wall actions
constructed from the standard Wilson kernel inherit its poor behavior.

In selecting a fermion action for thermodynamics, there should be no
excuse for deliberately building in poor continuum scaling.

\begin{table}
\begin{center}
\begin{tabular}{|c|c|c|c|}
\hline
action & $A_{2}/A_{0}$ & $A_{4}/A_{0}$ & $A_{6}/A_{0}$\tabularnewline
\hline
\hline
standard staggered & $248/147$ & $635/147$ & $3796/189$\tabularnewline
\hline
Naik & $0$ & $-1143/980$ & $-365/77$\tabularnewline
\hline
p4 & $0$ & $-1143/980$ & $73/2079$\tabularnewline
\hline
\hline
standard Wilson & $248/147$ & $635/147$ & $13351/8316$\tabularnewline
\hline
hypercube & $-0.242381$ & $0.114366$ & $-0.0436614$\tabularnewline
\hline
\hline
overlap/ & $248/147$ & $635/147$ & $3796/189$\tabularnewline
domain wall &~&~&\tabularnewline
\hline
\end{tabular}
\caption{Continuum limit scaling behavior of free massless quarks in
  various lattice formulations, based on an expansion
  (\protect\ref{eq:Pexpansion}) of the pressure in powers of
  $1/N_\tau^2$ from \protect\cite{Hegde:2008nx}.  Shown are ratios of
  the expansion coefficients to the ideal, leading Stefan-Boltzmann
  coefficient.  A small ratio indicates good scaling.
\label{tab:SB}
}
\end{center}
\end{table}

\begin{figure}
  \begin{tabular}{cc}
   \includegraphics[width=0.5\textwidth]{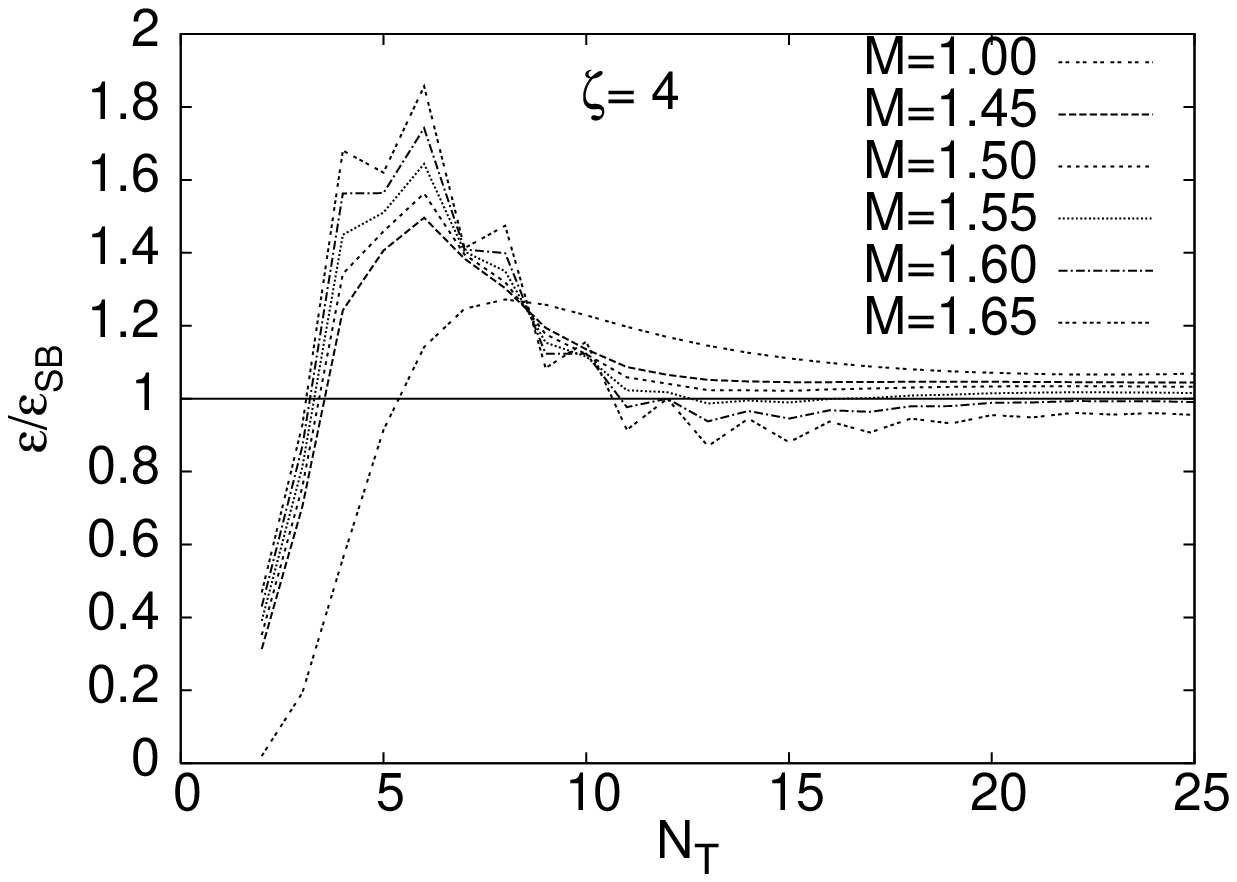}
    &
   \includegraphics[width=0.5\textwidth]{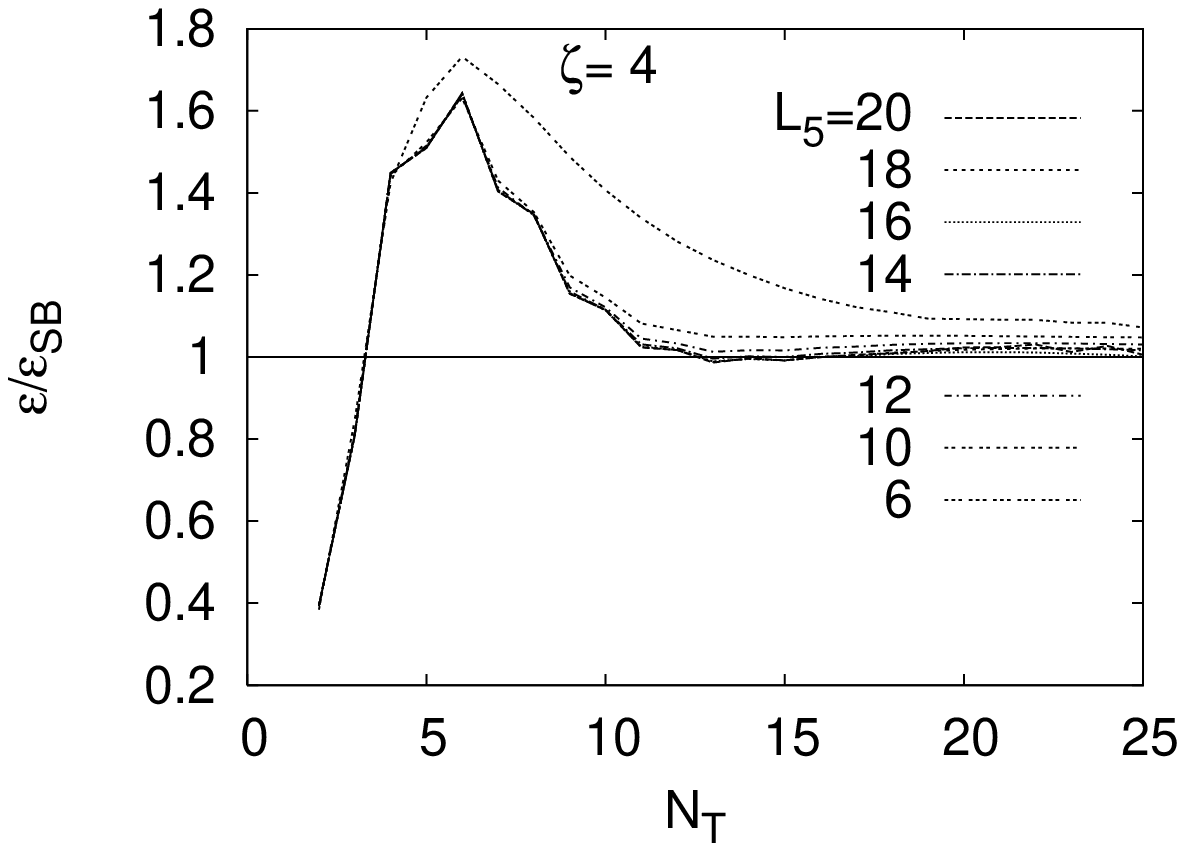} \\
   \end{tabular}
\caption{Deviation of the lattice free quark energy density from the
  Stefan-Boltzman continuum energy density as a function of $N_\tau$
  from \protect\cite{Gavai:2008uu}.  ($\zeta = N_s/N_\tau = 4$). Left
  panel: overlap for various mass shifts $M$.  Right panel: domain
  wall fermions for various $L_s$ at $M = 1.55$.
}
\label{fig:GS}
\end{figure}

Another recent study confirms pronounced cutoff problems with free
chiral fermion actions based on the standard Wilson kernel.  Gavai and
Sharma calculated the ratio of the lattice energy density to the
expected Stefan-Boltzmann value for overlap and domain wall fermions
\cite{Gavai:2008uu}.  Their results, reproduced in Fig.~\ref{fig:GS},
show slow continuum scaling and an oscillation related to negative
eigenvalues of the transfer matrix.

\begin{figure}
\begin{center}
\includegraphics[width=0.5\textwidth]{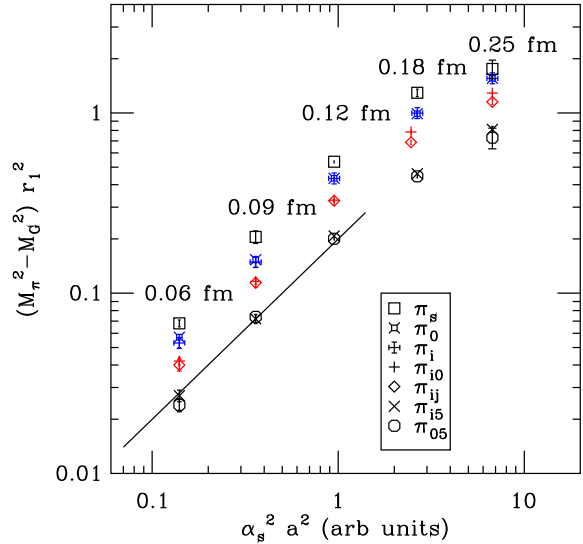}
\end{center}
\caption{Splitting of the pion taste multiplet \vs $\alpha_s^2 a^2$ for
  a wide range of lattice spacings. Splitting is measured as the
  difference of the squared masses of the member and the Goldstone
  member.  The plot symbols distinguish the members of the
  multiplet. The line is drawn with unit log-log slope to test
  proportionality to $\alpha_s^2 a^2$. }
\label{fig:taste}
\end{figure}

Staggered fermions have the awkward problem of extra ``taste'' degrees
of freedom.  The standard ``fourth root'' trick gives an approximately
correct counting of species, but hadrons in the statistical ensemble
still come in multiplets with a range of masses.  This is especially
visible in the pion spectrum.  The splitting of taste multiplets is
predicted to decrease as ${\cal O}(a^2\alpha_s^2)$ in terms of lattice
spacing and color fine structure constant
$\alpha_s$. Figure~\ref{fig:taste} confirms this trend for the pion
taste multiplet as a function of lattice spacing.

In judging how closely a simulation comes to the physical point,
devotees of staggered fermions may be tempted to focus on the lowest
member of the pion multiplet, \ie, the Goldstone boson pion.  While
this practice is correct when the Goldstone boson can be an isolated
external state, as in a zero temperature Green's function, in a
thermal ensemble, all members of the multiplet participate.  Thus it
is more appropriate in thermodynamics simulations to compare an
average multiplet mass, \eg, the rms pion mass with the physical pion
mass.  The physical point is reached only by reducing the lattice
spacing together with the light quark mass.  It is incorrect to claim
a thermodynamics calculation is done at a physical pion mass when the
rms mass is still much higher.

Is taste-symmetry breaking really a problem for thermodynamics?  It is
believed to be most dramatic for the pion and less noticeable for more
massive states \cite{Ishizuka:1993mt}.  One could argue that close to
the crossover temperature away from the critical point, so many
excited states participate, as in the resonating hadron gas model,
that pions do not matter much.  But if approach the critical point at
fixed lattice spacing, taste splitting is likely to have a strong
effect on the critical behavior: we may even get a chiral symmetry
restoring transition in the wrong universality class.  And certainly
at quite low temperatures where pions dominate the statistical
ensemble, taste splitting makes a difference.

\section{Phase Diagram and Determination of $T_c$}

\subsection{Current consensus}

\begin{figure}
\begin{center}
\includegraphics[width=0.5\textwidth]{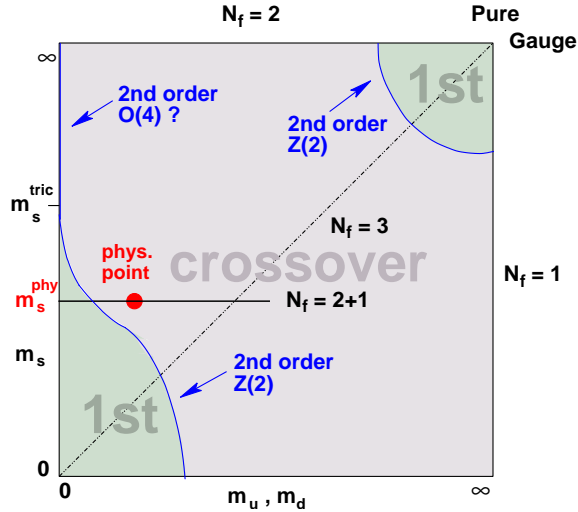} \\
\end{center}
\caption{Sketch from \protect\cite{Laermann:2003cv} of the phase
  diagram for QCD at zero baryon density in $2+1$ flavor QCD as a
  function of the light quark masses showing regions where a high
  temperature phase transition or crossover is expected. For a
  second-order phase transition, the universality class is shown.
  Whether the expected tricritical strange quark mass $m_s^{\rm tric}$
  is higher or lower than the physical strange quark mass $m_s^{\rm
    phys}$ is not yet firmly established.}
\label{fig:s_ud_plane}
\end{figure}

Is there a genuine phase transition separating a low temperature
confined phase with spontaneously broken chiral symmetry
from a high temperature deconfined phase with restored
chiral symmetry?  The answer depends on the number of quark flavors
and their masses.  Figure~\ref{fig:s_ud_plane} sketches the current
qualitative theoretical consensus for the case of 2+1 flavors of
quarks with masses $m_u = m_d$ and $m_s$ \cite{Laermann:2003cv}.  In
the upper right corner the quark masses are so large they play no role
in the statistical ensemble, and we enter the well-studied regime of a
first order confining/deconfining transition in pure Yang-Mills
theory.  At low temperature chiral models predict a first order
transition for degenerate masses, shown in the lower left cormer, and a
second order transition at large $m_s$ when $ m_u = m_d = 0$ (provided
the chiral anomaly does not vanish at the transition)
\cite{Pisarski:1983ms}.
\footnote{ In contradiction to this expectation, D'Elia, Di Giacomo,
  and Pica found indications of a first-order transition using an
  unimproved staggered fermion action and $N_\tau = 4$
  \cite{D'Elia:2005bv}.  It is important to check this conclusion with
  a more refined action.  In support of this expectation, Kogut and
  Sinclair have found a second order phase transition, but in the O(2)
  universality class, rather than O(4) \cite{Kogut:2006gt}.}  The
low-mass first order region is bounded by a critical line above which
the transition is only a crossover.

Figure~\ref{fig:s_ud_plane} is only qualitative.  To say whether there
is a phase transition at physical quark masses requires numerical
simulation.  The long-standing consensus has been that it is only a
crossover.  Aoki {\it et al.} have made a strong case for this
conclusion \cite{Aoki:2006we}.  Locating the actual critical line is
challenging, since it occurs at small quark masses and is quite
sensitive to cutoff effects \cite{Karsch:2001nf,Karsch:2003va}.
DeForcrand and Philipsen \cite{deForcrand:2006pv,deForcrand:Lat2008}
have done an impressively high statistics study that maps out the
phase boundary, but only with unimproved staggered fermions at $N_\tau
= 4$.  They have begun work at $N_\tau = 6$ \cite{deForcrand:Lat2008}.

\subsection{How precisely can we know $T_c$?}

So what is the crossover temperature $T_c$?  One must ask, first, for
what purpose do we need to know it?  If there is only a crossover, the
determination of $T_c$ is unquestionably imprecise, as the
Budapest-Wuppertal group has emphasized \cite{Aoki:2006br}.  For
phenomenology it should be good enough to determine the temperature
range over which an interesting quantity, such as the energy density
changes rapidly.  Each observable may give a somewhat different
answer.

For (presumably) unphysical quark masses for which a genuine phase
transition occurs, the transition temperature $T_c$ is unambiguous and
precision is achievable.  The observables used to locate it must,
obviously, have a sensible continuum limit, and they should expose the
critical behavior.

Finally, in determining the temperature there is a related question of
setting the lattice scale.  On coarse lattices, the result depends
strongly on the physical quantity used to set the scale.

With these preliminaries in mind we examine a variety of observables
that have been used to determine the crossover temperature.  First we
consider ``deconfinement-type'' observables.

\subsection{Deconfinement observables}

\subsubsection{Strange quark number susceptibility and equation of state}

The strange quark number susceptibility measures fluctuations in the
strange quark number:
\be
   \chi_s = \langle N_s^2 \rangle/(VT)
\ee
The energy density and pressure are also good indicators of the
progress of deconfinement.  

\begin{figure}
  \begin{tabular}{cc}
    \hspace*{-5mm}
    \includegraphics[width=0.6\textwidth]{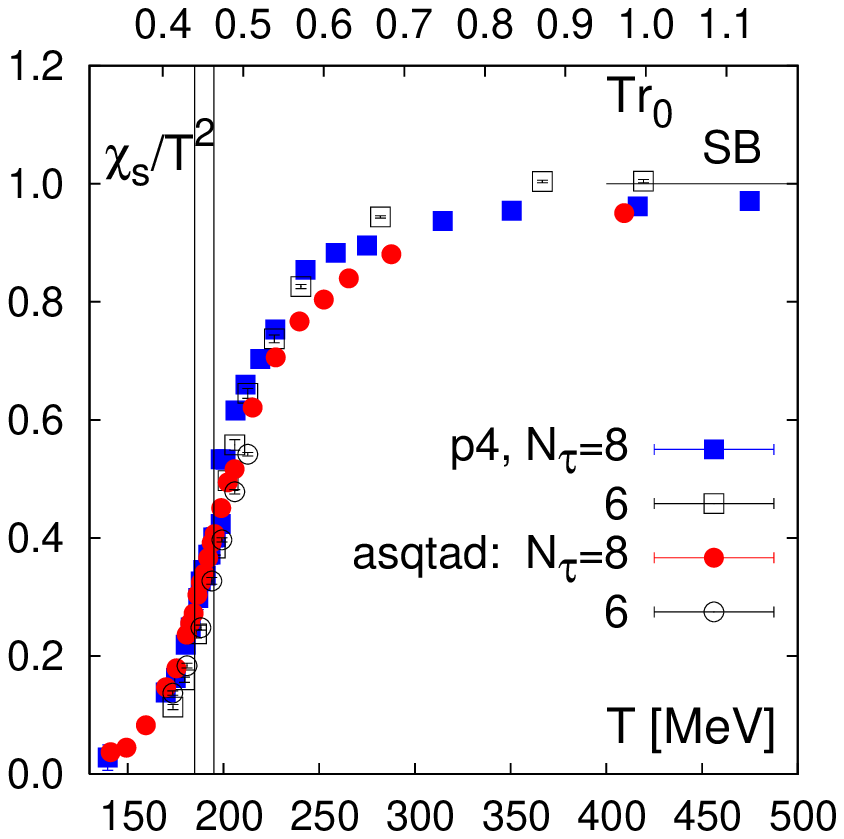}
    &
    \hspace*{-35mm}
    \includegraphics[width=0.6\textwidth]{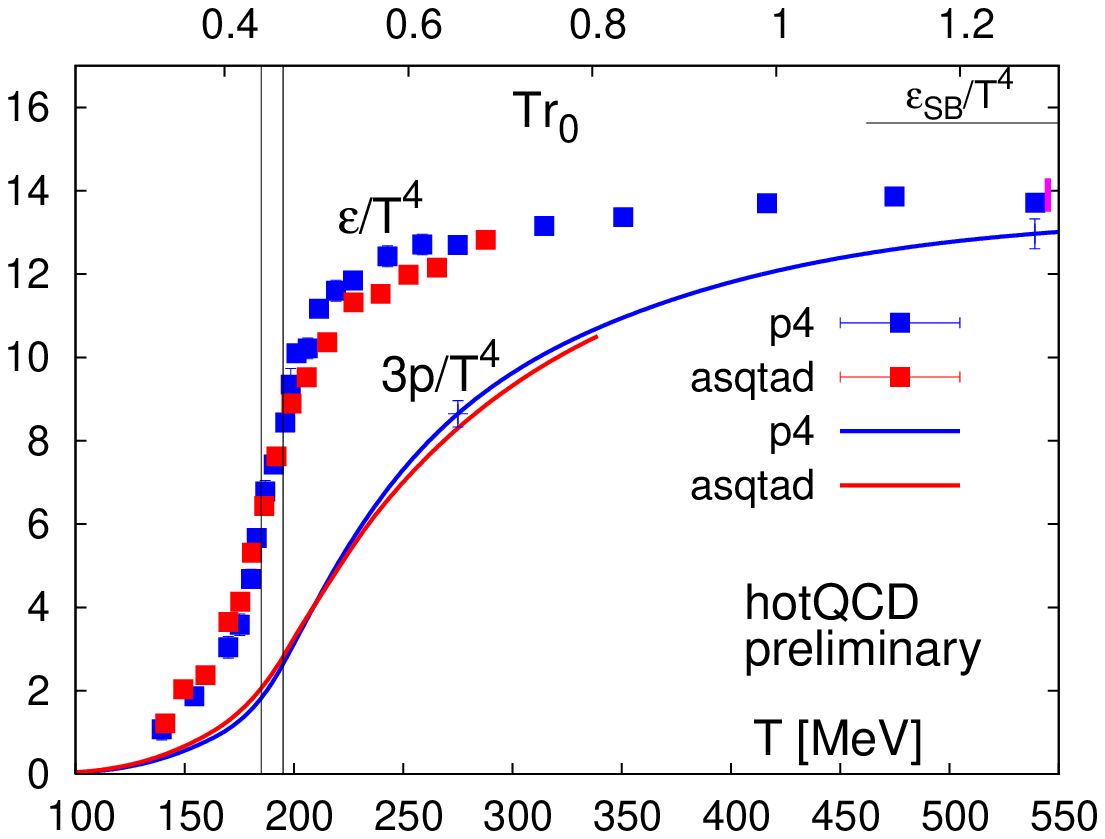}\\
   \end{tabular}
\caption{Left panel: Strange quark number susceptibility divided by
  the square of the temperature \vs temperature in MeV units (bottom
  scale) and $r_0$ units (top scale) for $N_\tau = 6$ and $8$.  Right
  panel: equation of state showing energy density and three times the
  pressure, both divided by the fourth power of the temperature \vs
  temperature for $N_\tau = 8$. Measurements are taken along a line of
  constant physics with $m_{ud} = 0.1 m_s$. These preliminary results
  are from a HotQCD study comparing p4fat3 and asqtad staggered
  fermion formulations \protect\cite{Gupta:Lat2008,DeTar:2007as}.  The
  blue error bars on the pressure curve indicate the size of the
  error.  The magenta bar shows a systematic error from setting the
  lower limit of the pressure integration.  The vertical bands here
  and in HotQCD figures below indicate a temperature range 185 -
  195 MeV and serve to facilitate comparison. }
\label{fig:strange_susc_and_e+3p}
\end{figure}

Recent results from the HotQCD collaboration are shown in 
Fig.~\ref{fig:strange_susc_and_e+3p} \cite{Gupta:Lat2008,DeTar:2007as}.

We note that the asqtad and p4fat3 results are in fair agreement. We
see that there are dramatic changes in both observables over the
temperature range 180 - 200 MeV.  The scale has been set through the
Sommer parameter $r_0$ (or $r_1$), which, in turn, is calibrated by
the measured splitting in bottomonium \cite{Aubin:2004wf}.  The rough
agreement between the $N_\tau = 6$ and $8$ gives some support for the
utility of this scale determination.

\subsubsection{Universal critical behavior}

To determine the crossover temperature more precisely, one may look
for the inflection point in the quark number susceptibility or a peak
in specific heat.  As we have remarked, the result is unambiguous only
for a genuine phase transition.  Both of these quantities are
derivatives of the free energy $f = -T \log Z$, which leads to a
unified treatment of their critical behavior \cite{Karsch:2007dt}.  At a
critical point, the free energy can be decomposed into singular and
analytic contributions.  The singular part scales according to
\cite{Karsch:2007dt,Hatta:2002sj}
\be
    f_s(T,\mu_q) = b^{-1}f_s(tb^{1/(2-\alpha)}) \sim t^{2 - \alpha} 
\ee
where from charge symmetry and analyticity, the scaling variable
depends on temperature and chemical potential through
\be
     t = \left|\frac{T-T_c}{T_c}\right| + c\left(\frac{\mu_q}{T_c}\right)^2.
\ee
For $O(4)$ the critical exponent is $\alpha \approx -0.25$.  This
relation can be used to predict singularities in quantities
expressible as derivatives of the free energy.  At $\mu_q = 0$ the
light quark number susceptibility is
\be
   \chi_\ell/T^2 \sim \frac{\partial^2 f_s}{\partial \mu^2} \sim t^{1-\alpha},
\ee
The singularity is weak and masked by analytic contributions.  Its
temperature derivative has a stronger singularity
\be
   \frac{\partial (\chi_\ell/T^2)}{\partial T} \sim t^{-\alpha}.
\ee
The same leading singularity is found in the specific heat
\be
    C_V \sim \frac{\partial^2 f_s}{\partial T^2} \sim t^{-\alpha} 
\ee
and in the quartic quark number susceptibility $c_4^q = (\langle
N_q^4\rangle - 3 \langle N_q^2 \rangle)$:
\be
    c_4^q \sim \frac{\partial^4 f_s}{\partial \mu_q^4} \sim t^{-\alpha} 
\ee

\subsubsection{Screening free energy of a static quark}

\begin{figure}
\begin{center}
\includegraphics[width=0.6\textwidth]{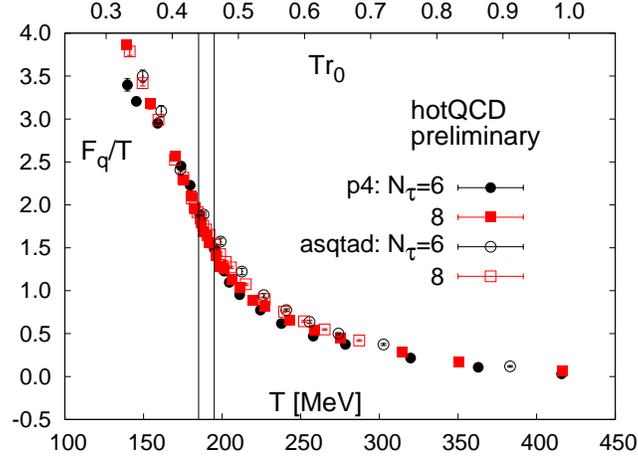}\\
\end{center}
\caption{Renormalized screening free energy of a static quark (from
  the renormalized Polyakov loop) \vs temperature in MeV units (bottom
  scale) and $r_0$ units (top scale) for $N_\tau = 6$ and $8$ from a
  HotQCD study comparing p4fat3 and asqtad staggered fermion
  formulations \protect\cite{Gupta:Lat2008,DeTar:2007as}. Measurements
  are taken along a line of constant physics with $m_{ud} = 0.1 m_s$.}
\label{fig:renormPloop}
\end{figure}

The Polyakov loop measures the screening free energy of a static quark
\cite{Bernard:1996zw}.  Thus it is a phenomenologically interesting
deconfinement-type observable.  It is not the derivative of the free
energy, so it might not reveal critical behavior.  It is customary to
renormalize it by removing a temperature-independent self-energy
\cite{Kaczmarek:2004gv}, leaving
\be
    F_q(T) = -T \log[P_{\rm renorm}(T)].
\ee
This quantity is shown in Fig.~\ref{fig:renormPloop}.  The related
susceptibility (essentially the variance in this quantity) is
sometimes used to locate the transition temperature, but in numerical
simulation the peak in this susceptibility weakens with increasing
$N_\tau$.

\subsection{Chiral observables}

Next, we consider observables usually associated with a chiral phase
transition.

\subsubsection{Chiral condensate}

The chiral condensate has an ultraviolet singularity for nonzero quark
mass and (at least for $N_f = 2$) an infrared chiral singularity at
zero quark mass $m_{ud}$:
\be
     \langle \bar \psi \psi \rangle(a,m_{ud},T) \sim \left\{
     \begin{array}{ll}
       c_{1/2}(a,T) \sqrt{m_{ud}} + c_1 m_{ud}/a^2 + {\rm analytic}
           & \quad \mbox{$T < T_c$} \\
       c_1 m_{ud}/a^2 + c_\delta m_{ud}^{1/\delta} + {\rm analytic} & \quad \mbox{$T = T_c$} \\
       c_1 m_{ud}/a^2 + {\rm analytic}
           & \quad \mbox{$T > T_c$}
     \end{array}
     \right. 
     \label{eq:pbp_sing}
\ee
The ultraviolet singularity appears in perturbation theory at the
one-quark-loop level.  It is temperature independent.  The infrared
singularity occurs in the chirally broken phase $T < T_c$
\cite{Karsch:2008ch}.  It is seen in chiral perturbation theory at the
one-pion-loop level.  A square root is the thermal analog of a chiral
log.  At $T_c$ we have the expected critical behavior.

\begin{figure}
\begin{center}
\hspace*{15mm}
\includegraphics[width=0.6\textwidth]{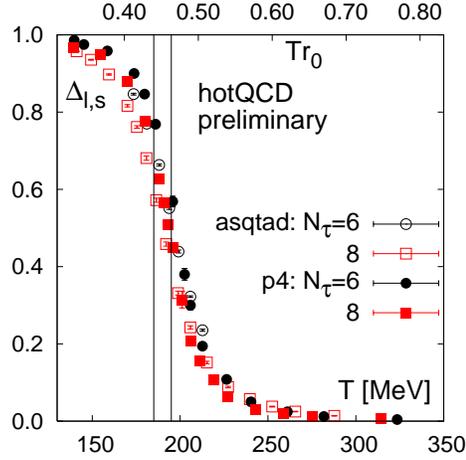}\\
\end{center}
\caption{Chiral condensate difference ratio \vs temperature in MeV
  units (bottom scale) and $r_0$ units (top scale) for $N_\tau = 6$
  and $8$ from a HotQCD study comparing p4fat3 and asqtad staggered
  fermion formulations
  \protect\cite{Gupta:Lat2008,DeTar:2007as}. Measurements are taken along
  a line of constant physics with $m_{ud} = 0.1 m_s$.}
\label{fig:diff}
\end{figure}

The ultraviolet singularity in this quantity poses a problem for the
continuum limit.  Since the condensates for all flavors have this
singularity, taking an the appropriate linear combination of light and
strange quark condensates removes it, and dividing by the
zero-temperature value removes the multiplicative renormalization factor
\cite{Karsch:2007dt}:
\be
     D_{\ell,s}(T) = \langle \bar \psi \psi \rangle|_\ell 
         - \frac{m_\ell}{m_s} \langle \psi \psi \rangle|_s
     \ \ \ \ \ \ \ \ 
     \Delta_{\ell,s}(T) = D_{\ell,s}(T)/D_{\ell,s}(T=0)
\ee
The difference ratio $\Delta_{\ell,s}$ is shown in
Fig.~\ref{fig:diff}.  We see a dramatic drop in this quantity over
approximately the same temperature range over which we saw a rapid
rise in Fig.~\ref{fig:strange_susc_and_e+3p}.

\subsubsection{Chiral susceptibility}

The chiral susceptibility measures fluctuations in the chiral
condensate.  Loosely speaking it is
\be
 \chi = 
   \partial \left\langle \bar \psi \psi \right \rangle(a,m,T)/\partial m
  \label{eq:chiral_susc}
\ee
where it is customary to distinguish the connected and disconnected
contributions in terms of the Dirac matrix $M$:
\bea
 \chi_{\rm disc} &=& 
   \frac{T}{V}\left[
   \left\langle (\Tr M^{-1})^2\right\rangle - 
           \left\langle \Tr M^{-1} \right\rangle^2\right] \nonumber \\
  \chi_{\rm conn} &=& -\frac{T}{V}\left\langle \Tr M^{-2} \right\rangle.
\eea
Tradition holds that a peak in $\chi_{\rm disc}$ marks the crossover,
but see the discussion of possible distortions below.

\begin{figure}
  \begin{tabular}{cc}
  \includegraphics[width=0.6\textwidth]{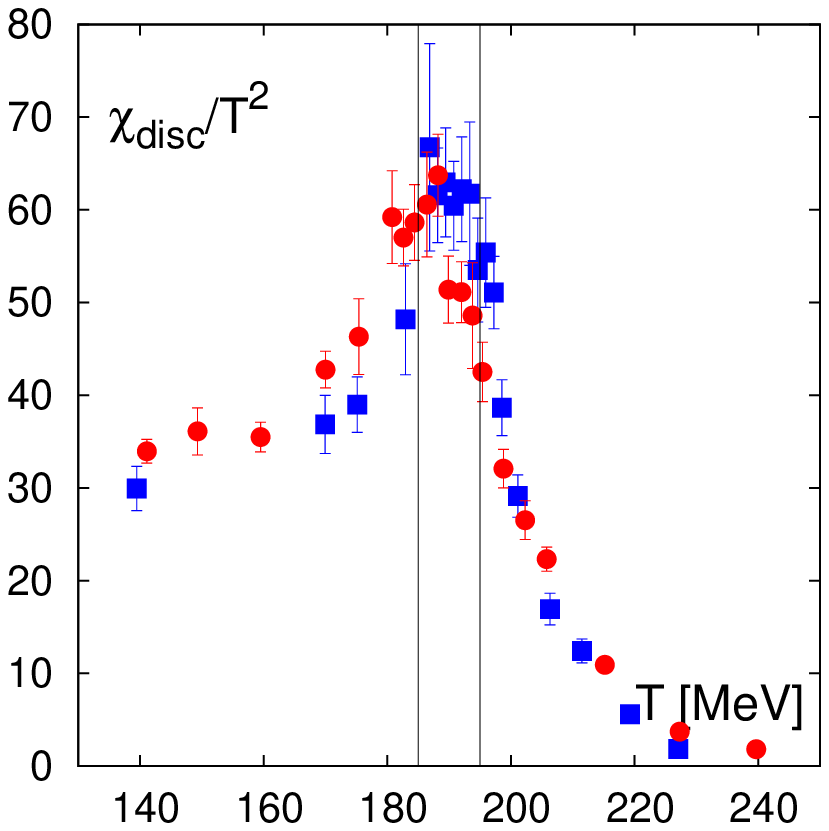}
  &
  \hspace*{-25mm}
  \includegraphics[width=0.6\textwidth]{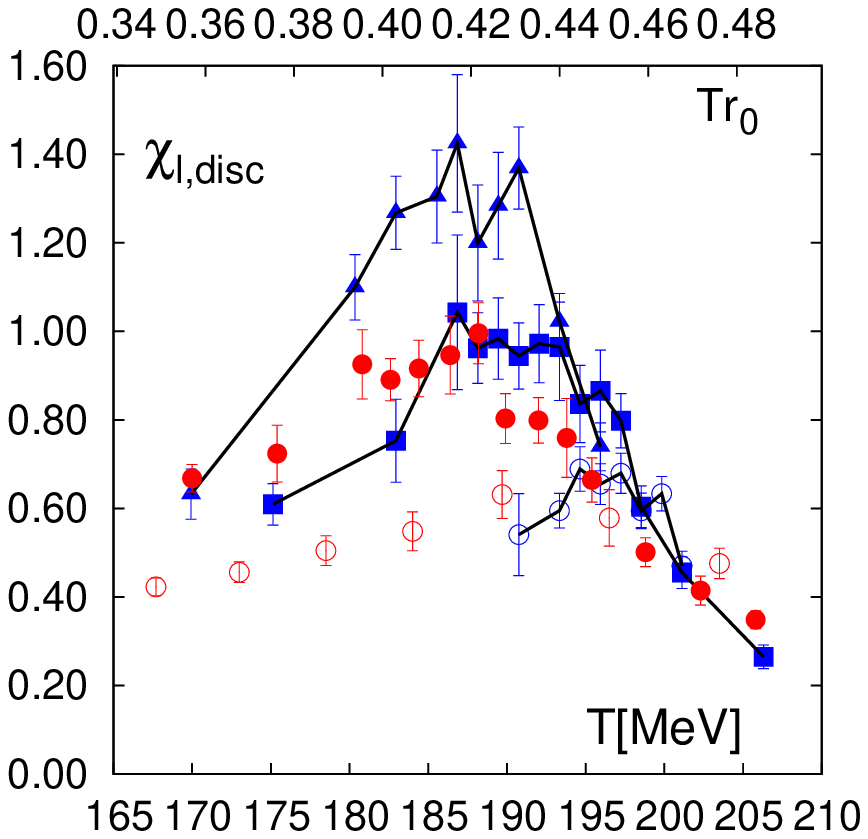}\\
   \end{tabular}
\caption{Left panel: Disconnected light quark susceptibility \vs
  temperature in MeV units (bottom scale).  Right panel: closeup of
  the peak region. Lines merely connect the points. Red symbols, asqtad
  fermions.  Blue symbols, p4fat3.  Filled squares and circles are
  along a line of constant physics with $m_{ud} = 0.1 m_s$. Open
  circles, with $m_{ud} = 0.2 m_s$, filled triangles with $m_{ud} =
  0.05 m_s$.  All results are HotQCD preliminary
  \protect\cite{Gupta:Lat2008,Soeldner:Lat2008,DeTar:2007as}.}
\label{fig:asqtad_p4_chiral_sus}
\end{figure}

Figure~\ref{fig:asqtad_p4_chiral_sus} shows recent results for the
disconnected susceptibility.  The peak occurs in roughly the same
temperature range as the dramatic changes seen in the previous
observables in Figs.~\ref{fig:strange_susc_and_e+3p} and
\ref{fig:diff}.  The peak height increases as expected as the light
quark mass is decreased.

\begin{figure}
\begin{center}
  \includegraphics[width=0.7\textwidth,clip=on]{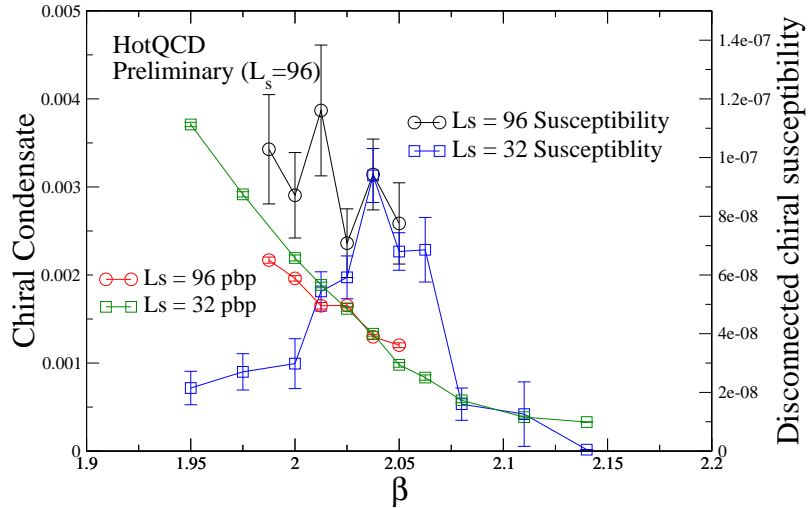} \\
\end{center}
\caption{Chiral condensate (left scale) and disconnected chiral
  susceptibility (right scale) \vs the gauge coupling parameter
  $\beta$.  Results are from a HotQCD study of $2+1$ flavor domain
  wall thermodynamics at $N_\tau = 8$ for two choices of $L_s$
  \cite{Cheng:Lat2008}. Measurements are taken with light and strange
  quark masses fixed in lattice units ($am_{ud} = 0.003$ and $am_s =
  0.37$).}
\label{fig:DW_chiral_sus}
\end{figure}

The first exploratory $N_\tau = 8$ results for the domain wall action
\cite{Cheng:Lat2008} are shown in Fig.~\ref{fig:DW_chiral_sus}.
Because domain wall calculations are vastly more expensive, the domain
wall effort is certainly not as advanced as other efforts.  The $L_s =
96$ study was undertaken to assure a small residual quark mass over
the range of couplings shown.

The derivative in Eq (\ref{eq:chiral_susc}) generates the correlator
of the relevant chiral condensates, integrated over the space-time
volume:
\bea
  \chi &=& C(p=0,T) = \int d^4x \, C(x,T) \nonumber \\
  C(x,T) &=& \left\langle \bar\psi\psi(x) \bar\psi\psi(0)\right\rangle
  \label{eq:susc_correl}
\eea
To be more precise, the derivative in the definition of the
susceptibility (\ref{eq:chiral_susc}) can involve any of the flavor
condensates and any of the quark masses.  For the light quark
condensates, it is useful to distinguish the isosinglet and isotriplet
chiral condensates, according to the isospin content of the operators
in the correlator.  These quantities are linear combinations of the
disconnected and connected susceptibilities, namely, $\chi_{\rm sing}
= \chi_{\rm disc} + 2 \chi_{\rm conn}$ and $\chi_{\rm trip} = 2
\chi_{\rm conn}$.

The ultraviolet and infrared singularities of the isosinglet chiral
susceptibility can be easily inferred from Eq (\ref{eq:pbp_sing})
\cite{Karsch:2008ch}:
\be
     \chi_{\rm sing} \sim \left\{
     \begin{array}{ll}
       c_1/a^2 + c_{1/2}(a,T)/(2\sqrt{m}) + {\rm analytic}
           & \quad \mbox{$T < T_c$} \\
       c_1/a^2 + c_\delta m^{1/\delta-1} + {\rm analytic} & \quad \mbox{$T = T_c$} \\
       c_1/a^2 + {\rm analytic}
           & \quad \mbox{$T > T_c$}
     \end{array}
     \right.
\label{eq:chiral_susc_singularities}
\ee
We see that it also suffers from an ultraviolet divergence, which
makes it increasingly noisy as the lattice spacing is decreased.  The
infrared singularity at zero quark mass in the chirally broken phase
($T < T_c$) arises from the vanishing of the pion mass in that limit.
For this reason at small quark mass we should expect not only a peak
at the transition point, but we should expect the susceptibility to
grow at lower temperatures as well \cite{Karsch:2008ch}.

To remove the ultraviolet singularity the Budapest/Wuppertal group
subtracts the zero temperature value and multiplies by the square of
the quark mass to cancel multiplicative renormalization factors:
\be
    m_q^2[\chi(m_q,T) - \chi(m_q,0)]/T^4
\ee
Compared with the uncorrected susceptibility, along a line of constant
physics away from the chiral limit, this definition tends to shift a
peak to lower $T$ because $m_q^2/T^2$ decreases with increasing $T$.
If instead of a peak, there is a shoulder, it might induce a peak.

\begin{figure}[t]
\begin{center}
\includegraphics[width=0.4\textwidth]{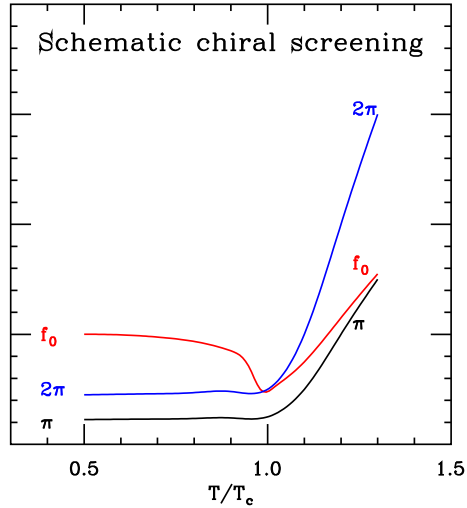}
\end{center}
\caption{Sketch of the expected behavior of the light screening spectrum
  for the $\pi$, $f_0$ and $2 \pi$ threshold \vs temperature in units
  of the crossover temperature $T_c$.}
\label{fig:sketch_screen}
\end{figure}

\subsubsection{Screening masses as indicators of the transition}

From Eq (\ref{eq:susc_correl}) we see that a spectral component of
mass $M(T)$ and weight $\rho(M,T)$ in the correlator contributes
$\rho(M,T)/M(T)^2$, which is singular when $M(T)$ vanishes.  Since the
correlator is integrated over imaginary time, we can analyze the
spatial dependence of the zero-Matsubara-frequency correlator to
determine its spectrum.  The masses in that case are ``screening
masses'' \cite{DeTar:1987ar}.  Figure~\ref{fig:sketch_screen} sketches
a possible scenario for the temperature dependence of the low spectral
components of the isosinglet chiral condensate.  In the chiral limit
the $f_0$ must be degenerate with the pion for $T > T_c$, so for a
continuous transition, it must drop to zero there.  As the light quark
mass is decreased, the two-pion threshold also vanishes, leading to an
infinite chiral susceptibility for $T < T_c$ as well, as reflected in
Eq (\ref{eq:chiral_susc_singularities}).  Close to the chiral limit,
instead of a peak marking the transition, one might expect a cliff.

\begin{figure}
  \begin{center}
  \includegraphics[width=0.7\textwidth]{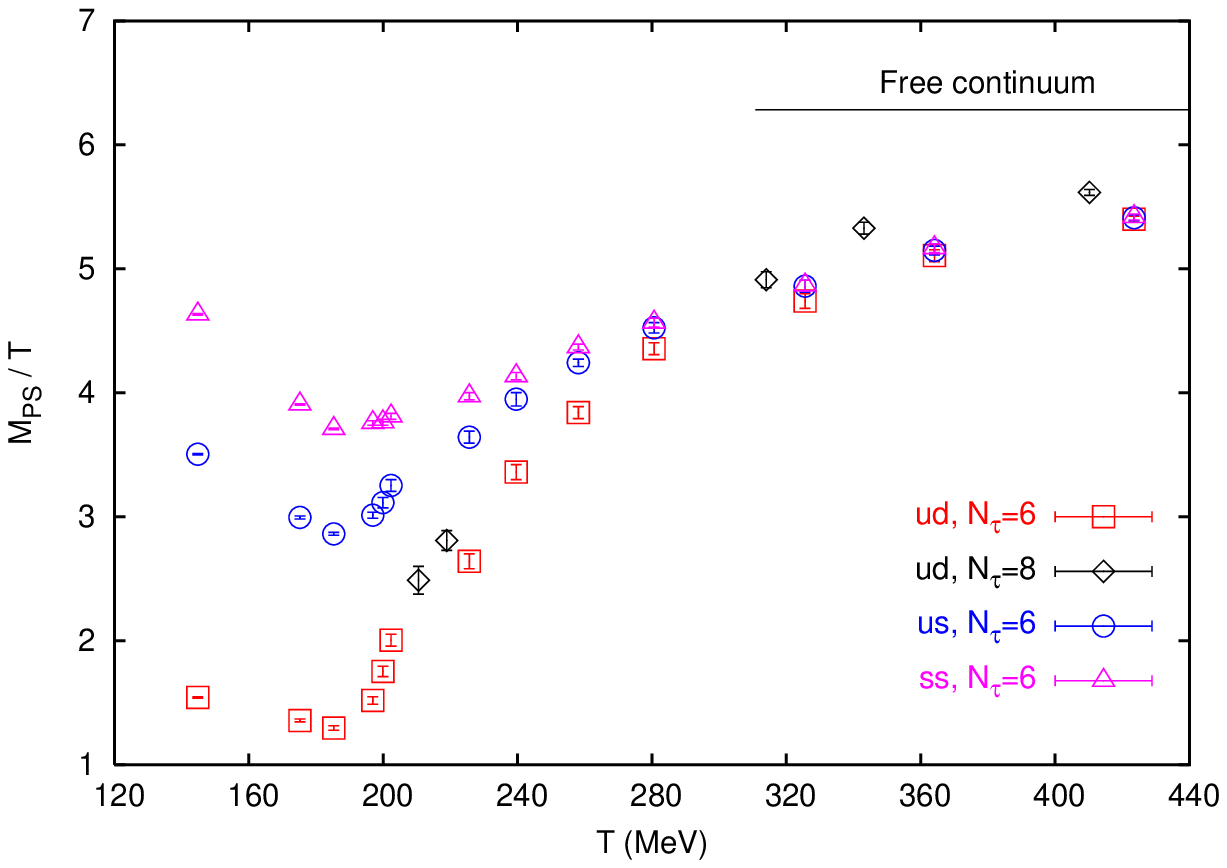} \\
  \includegraphics[width=0.7\textwidth]{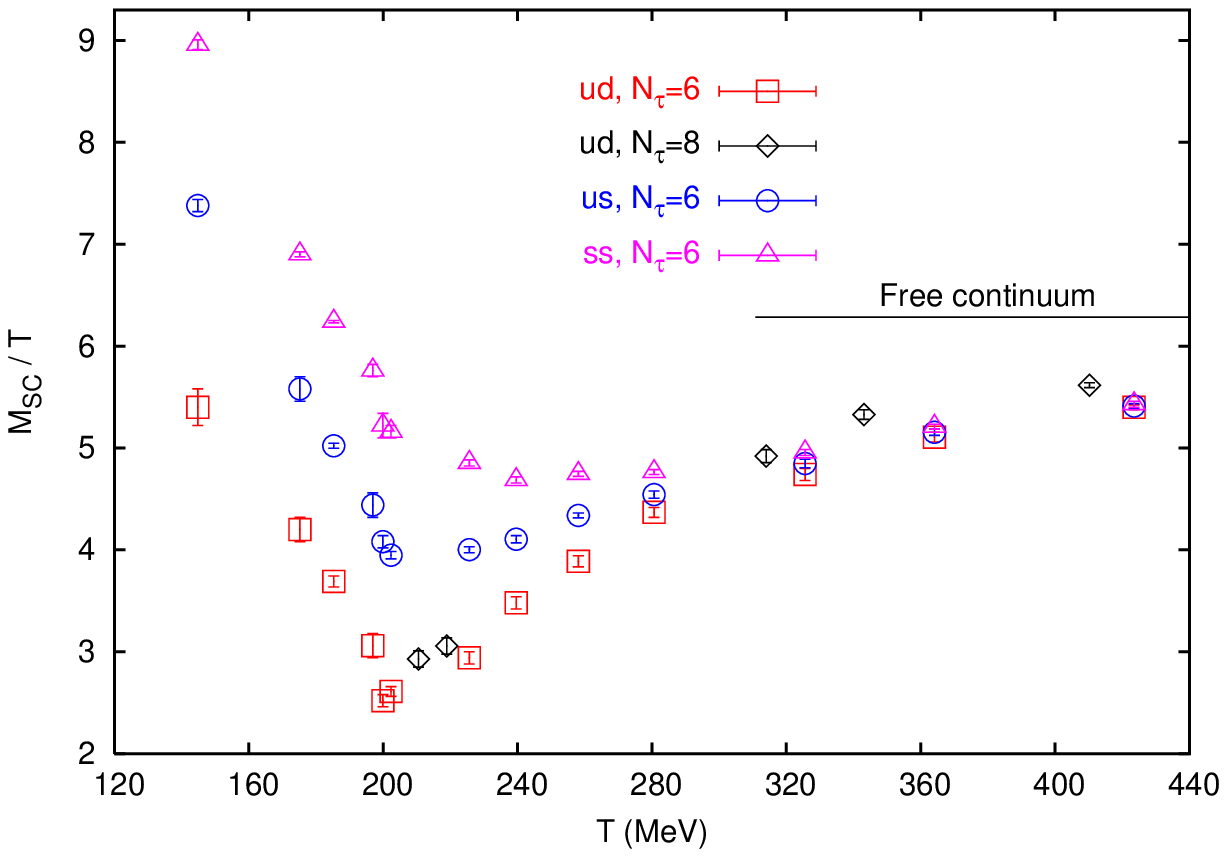} \\
  \end{center}
\caption{Screening masses for the pseudoscalar channel (upper panel)
  and scalar channel (lower panel) \vs temperature in a dynamical
  $2+1$ flavor simulation with p4fat3 staggered fermions
  \cite{Cheng:2007jq}.  Measurements were taken along lines of
  constant physics with $m_\pi \sim 220$ MeV, $m_K = 500$ MeV and $N_\tau
  = 6$ and $8$ \cite{Laermann:Lat2008}. }
\label{fig:RBCBscreen}
\end{figure}

At this conference on behalf of the RBC/Bielefeld collaboration
Laermann reported new measurements of screening masses in the scalar
and vector isotriplet channels \cite{Laermann:Lat2008}. Some of their
results are shown in Fig.~\ref{fig:RBCBscreen}.  We see that the $\pi$
shows the behavior sketched in Fig.~\ref{fig:sketch_screen}.  Results
for the $f_0$ are not available, but the $a_0$ shows a steep dip at
the crossover.

As we see, screening masses may prove to be a useful indicator of the
chiral transition, since they do not suffer from ultraviolet or chiral
divergences, and they require no renormalization.

\subsection{Scale determination}

To quote the transition temperature $T_c$ in physical units requires a
scale determination.  The Budapest/Wuppertal group favors setting the
lattice scale with $f_K$, whereas MILC and RBC/Bielefeld use the Sommer
parameter $r_0$ or the related $r_1$.  At current typical lattice
spacings and quark masses in staggered fermion simulations the $f_K$
scale results in a 10 to 20\% lower temperature in MeV than the $r_0$
scale.  This discrepancy vanishes at the physical quark mass and in
the continuum.  We should choose the scale so that the crossover
temperature scales well.  Of course, given the ambiguities in
determining the crossover temperature, even in lattice units, that is
an imprecise condition.  The deconfinement-type variables are more
useful to phenomenology.  For them the $r_0$ scale seems to give
reasonably consistent results as the lattice spacing is decreased, as
we can see from Figs.~\ref{fig:strange_susc_and_e+3p} and
\ref{fig:diff} and Aoki {\it et al.}  \cite{Aoki:2006br} (Fig.~4).
Thus there appears to be no reason to abandon the $r_0$ scale for now.

In the past two years there have been some seemingly contradictory
estimates of the transition temperature.  Aoki {\it et al.} reported
that at physical quark masses in the continuum, $T_c = 151(3)(3)$ MeV
from a peak in the chiral susceptibility and $T_c = 175(2)(4)$ from
the inflection point in the quark number susceptibility and the
Polyakov loop variable.  These are to be compared with an older result
from the MILC collaboration $169(12)(4)$ MeV based on the chiral
susceptibility \cite{Bernard:2004je} and a more recent determination
of $192(7)(4)$ MeV by the Bielefeld/RBC group based on a combination
of chiral and deconfining observables \cite{Cheng:2006qk}.  The last
two groups used the $r_0$ (or $r_1$) scale.  The Budapest/Wuppertal
group has carefully listed sources of the discrepancy, which include
ambiguities in locating the crossover, their preferred renormalization
of the chiral susceptibility, and their preference for the $f_K$
scale.  This year we can add to the list the possibility that the
chiral susceptibility develops an asymmetric peak or shoulder, which
would be even more sensitive to the renormalization procedure and
should not be modeled by a parabola.  A combination of these effects
could certainly account for the discrepancy.

\section{Equation of state}

The equation of state is fundamental to hydrodynamic calculations of
the expansion of hot hadronic matter.

\subsection{Standard integral method}

The currently popular method for calculating the equation of state
begins with the lattice-thermodynamic identity at fixed $N_\tau$,
which expresses the trace of the energy momentum tensor or
``interaction measure'' $I$ in terms of the derivative of the log of
the partition function $Z$:
\be
    I = \varepsilon - 3p = -\frac{T}{V} \frac{d \ln Z}{d \ln a}.
\label{eq:int_meas}
\ee
The derivative with respect to lattice spacing is taken with fixed
output hadron masses.  Thus it involves the derivative of the bare
lattice parameters, \ie, the gauge coupling $g$, quark masses, and for
some actions the tadpole coefficient $u_0$, with respect to the cutoff
scale and the expectation values of the action operators.  All of
these nonperturbative quantities are readily calculated in lattice
simulations.

We normalize the energy and pressure to zero at zero temperature.
This eliminates an ultraviolet divergence of order $1/a^4$.  This is
done by subtracting the zero temperature quantity, calculated with the
same bare parameters.  Because we are subtracting two ultraviolet
singular quantities, we must increase the simulation sample size
dramatically as the lattice spacing is decreased.  The calculational
cost thus grows steeply as we approach the continuum limit.  In what
follows we will assume this subtraction has been done for all
thermodynamic quantities.

A second identity becomes
\be
   \frac{p}{T} = \left.\frac{\partial \ln Z}{\partial V}\right|_T  
   \rightarrow (\ln Z)/V
\label{eq:pressure}
\ee
in the thermodynamic limit for which $\ln Z \propto V$. Finite-size
deviations from this limit could produce deviations from the
Stefan-Boltzmann law \cite{Gliozzi:2007jh,Panero:Lat2008}. Such finite
size effects may be important at ultrahigh $T$, where we would like to
compare with perturbation theory \cite{Endrodi:2007tq}.

Putting Eqs (\ref{eq:int_meas}) and (\ref{eq:pressure}) together gives
the the pressure as the integral of the interaction measure from
coarse to fine lattice spacing, \ie, low to high temperature:
\be
    \left.\frac{Vp}{T}\right|_a -\left.\frac{Vp}{T}\right|_{a_0}  
          = -\int_{\ln a_0}^{\ln a}\frac{V^\prime}{T^\prime}
                (\varepsilon^\prime - 3p^\prime)\, d\ln a^\prime.
\ee
If the lower limit is sufficiently low in temperature, the pressure is
zero.  With pressure and interaction measure in hand we immediately
get the energy density and entropy.

\begin{figure}[t]
  \begin{tabular}{ccc}
   \hspace*{-15mm}
  \includegraphics[width=0.5\textwidth]{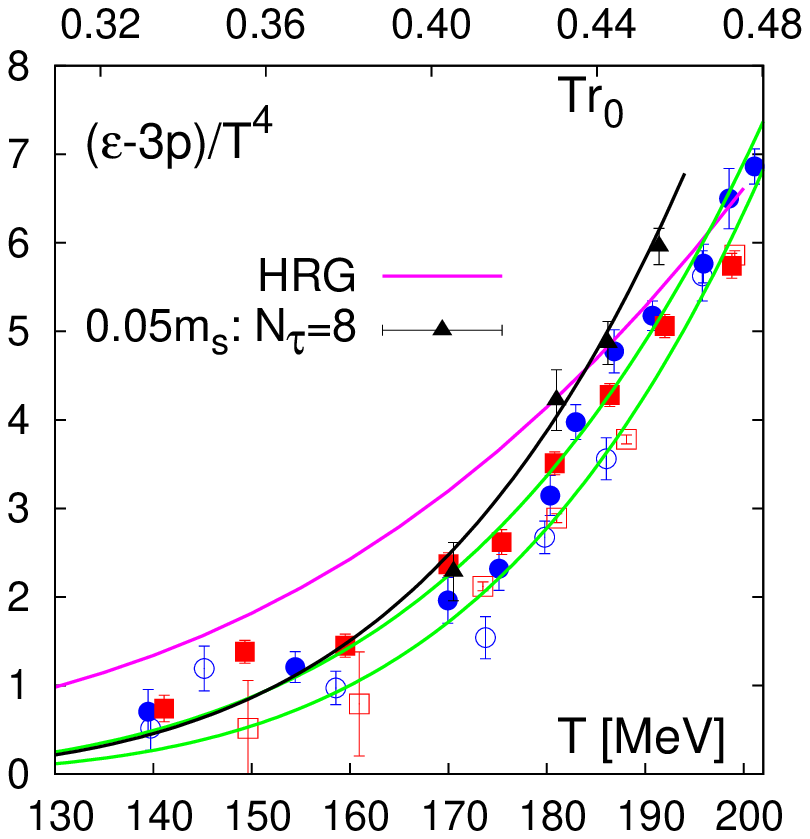}
 &
  \hspace*{-20mm}
  \includegraphics[width=0.5\textwidth]{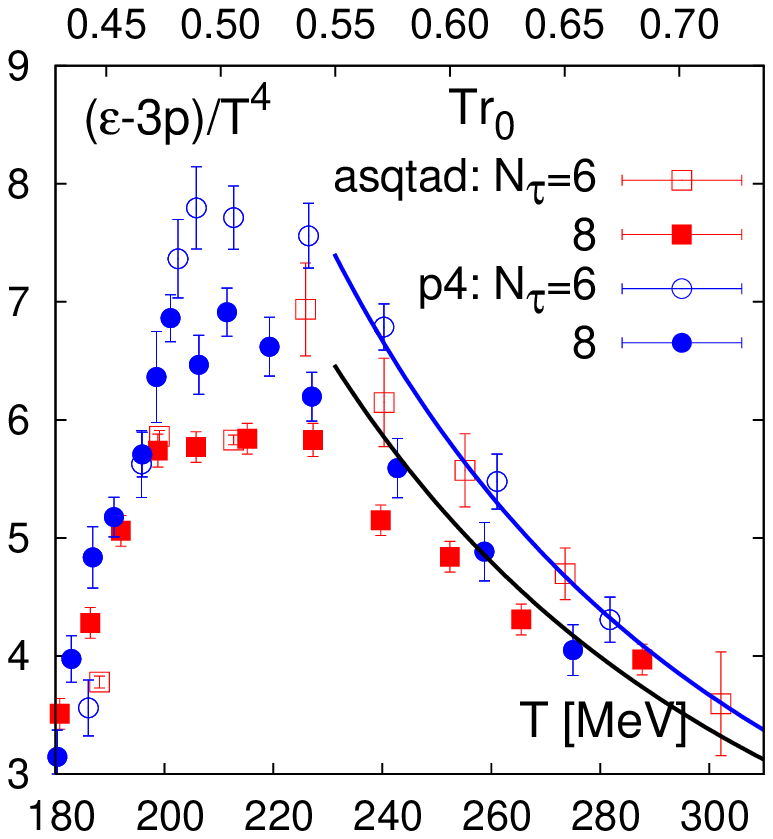}
 &
  \hspace*{-30mm}
  \includegraphics[width=0.5\textwidth]{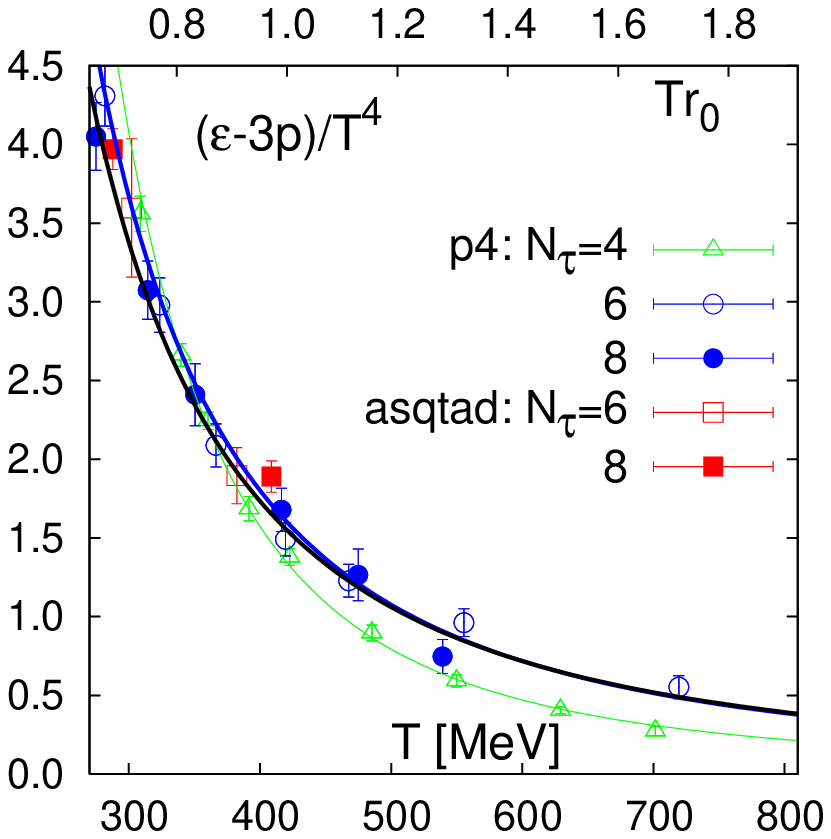} \\
 \end{tabular}
\caption{Details of the dependence of the interaction measure on
  temperature in MeV units (bottom scale) and $r_0$ units (top scale)
  for three temperature ranges left to right: low, mid, and high, for
  $N_\tau = 6$ and $8$ from a HotQCD study comparing p4fat3 and asqtad
  staggered fermion formulations
  \protect\cite{Gupta:Lat2008,Soeldner:Lat2008,DeTar:2007as}. Measurements
  in most cases are taken along a line of constant physics with
  $m_{ud} = 0.1 m_s$. Results in the high temperature range at $N_\tau
  = 4$ are from \protect\cite{Cheng:2007jq}. In the low temperature
  range the magenta curve is the prediction of a hadron resonance gas
  model.  The other curves in that range are spline fits to the data.
  The curves in the high temperature range are fits to a quadratic in
  $1/T^2$. }
\label{fig:e-3p}
\end{figure}

\begin{figure}
\begin{center}
\includegraphics[width=0.6\textwidth]{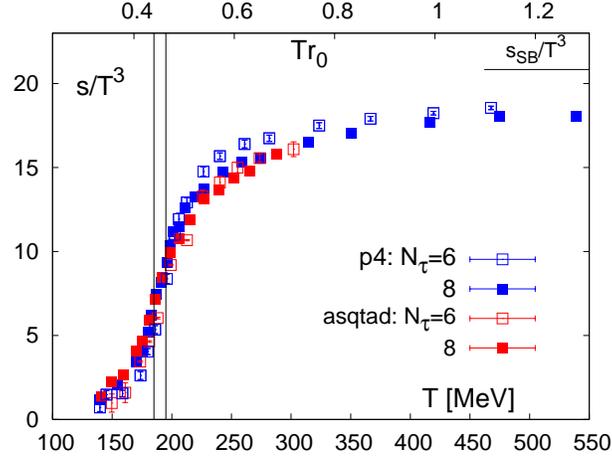} \\
\end{center}
\caption{Entropy density divided by the third power of the temperature
  \vs temperature in MeV units (bottom scale) and $r_0$ units (top
  scale).  for $N_\tau = 6$ and $8$ from a HotQCD study comparing
  p4fat3 and asqtad staggered fermion formulations
  \protect\cite{Gupta:Lat2008,DeTar:2007as}.  Measurements in most
  cases are taken along a line of constant physics with $m_{ud} = 0.1
  m_s$.}
\label{fig:entropy}
\end{figure}

Recent results for the interaction measure for $N_\tau = 6$ and $8$
are shown in Fig.~\ref{fig:e-3p}.  The resulting equation of state and
pressure were shown in Fig.~\ref{fig:strange_susc_and_e+3p} and the
entropy density is shown in Fig.~\ref{fig:entropy}.  Whether there is
a statistically significant disagreement between the asqtad and p4fat3
results in the central region remains to be determined after further
data are accumulated.  In the high temperature range $T \in [250,700]$
MeV, the results can be fit to
\be
  (\varepsilon - 3p)/T^4 = b/T^2 + c/T^4
\ee
There seems to be no need yet to include perturbative $1/\log T$
terms in the fit coefficients $b$ and $c$.  The plot in the low
temperature range compares the lattice result with predictions of a
hadron resonance gas model.  Since the lattice calculations include
cutoff effects, it would be premature to draw conclusions based on a
disagreement at this level.

\subsection{New $T$ integral method}

In the past year the WHOT collaboration introduced a new method in
which the integral over lattice spacing at fixed $N_\tau$ is replaced
by an approximate integral over $N_\tau$ at fixed lattice spacing (\ie,
fixed bare lattice parameters) \cite{Umeda:Lat2008,Umeda:2008bd}.

The method starts from an alternative form of Eq (\ref{eq:int_meas})
with Eq (\ref{eq:pressure}):
\be
  I/T^4 = \frac{d(p/T^4)}{d \ln T}.
\label{eq:int_meas2}
\ee

The pressure is then computed by integrating the interaction measure
with respect to $\ln T$ or equivalently $\ln(N_\tau)$.  Since $N_\tau$
is an integer, to reduce discretization errors in the sampling of the
integrand, one must reduce the temporal lattice spacing $a_t$.  An
anisotropic lattice helps.

\begin{figure}
\begin{center}
\includegraphics[width=0.5\textwidth]{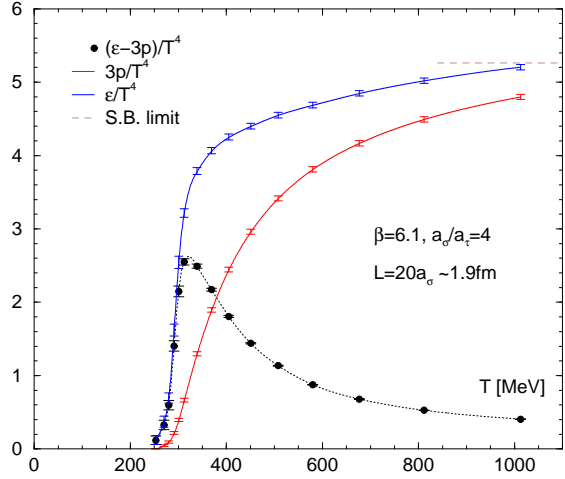} \\
\end{center}
\caption{Equation of state (interaction measure, energy density and
  pressure) for pure Yang-Mills theory, obtained using the $T$
  integral method at fixed lattice spacing $a_\sigma = 0.097$ fm and
  aspect ratio $a_\tau/a_\sigma = 4$
  \cite{Umeda:Lat2008,Umeda:2008bd}. }
\label{fig:tumeda-fig1}
\end{figure}

With this potentially computationally cheaper method the zero
temperature subtraction is common to all $N_\tau$, and with bare
lattice parameters fixed, one necessarily works along lines of
constant physics. Figure~\ref{fig:tumeda-fig1} shows the result of a
test calculation done for pure SU(3) Yang-Mills theory.

\section{Plasma Structure}

In addition to the phase diagram and equation of state, lattice
simulations provide information about the structure of hot hadronic
matter.  I mention two recent developments: a new effort to determine
the shear and bulk viscosity over a range of temperatures and a new
study of the spatial string tension.

\subsection{Transport Coefficients}

Analysis of RHIC heavy ion collisions suggests that high temperature
hadronic matter is an exceptionally good fluid.  To confirm this
hydrodynamical characterization requires computing the transport
coefficients, namely the shear ($\eta$) and bulk ($\zeta$) viscosities.
They are obtained from correlators of the energy-momentum tensor at
temperature $T$
\be
  C(x_0,{\bf x},T) =  
   \left\langle T_{\mu\nu}(x_0,{\bf x})T_{\rho\sigma}(0).
            \right\rangle
\ee
We need its spectral function $\rho$, which we obtain from the Kubo
formula
\be
   C(x_0,{\bf q},T) =  \int_0^\infty d\omega \, \rho(\omega,{\bf q}, T)
            \frac{\cosh \omega(x_0 - 1/2T)}{\sinh(\omega/2T)}.
\ee
The transport coefficients are obtained from the low-frequency behavior
of the spectral function
\be
    \eta(T) = \pi \lim_{\omega\rightarrow 0} 
          \frac{\rho_{12,12}(\omega, 0, T)}{\omega} 
    \ \ \ \ \ \ \ \ \ \ \ \ 
    \zeta(T) = \frac{\pi}{9} \lim_{\omega\rightarrow 0} 
          \frac{\rho_{ii,jj}(\omega, 0, T)}{\omega}
\ee
This has been a well known challenging problem since it was first
attempted by Karsch and Wyld \cite{Karsch:1986cq}.  The correlator is
noisy, requiring high statistics.  Going from a Euclidean correlator
$C(x_0)$ to $\rho(\omega)$ is a very difficult inverse problem.
Because of time-reflection symmetry, a simulation at $N_\tau = 8$ has
only five, typically noisy, independent values.

Possible remedies include (1) assuming a functional form for $\rho$
and fitting its parameters, (2) decreasing the time interval $a_t$,
allowing a larger $N_\tau$, and (3) adding further constraints on
$\rho$, such as maximum entropy.

Meyer \cite{Meyer:Lat2008,Meyer:2007dy,Meyer:2007ic} has done a new
high statistics calculation in pure Yang-Mills theory and uses a
paramerization of the spectral function in terms of an optimized basis
set that folds in appropriate perturbative behavior.  For the ratio of
shear viscosity to entropy density, he finds $\eta/s = 0.134(33)$ at
$1.65 T_c$ where perturbation theory gives 0.8, and for the ratio of
bulk viscosity to entropy density, $\zeta/s < 0.15$ at $1.65 T_c$ and
$\zeta/s < 0.015$ at $3.2 T_c$.

\subsection{Spatial string tension}

Despite its popular characterization as deconfined, high temperature
hadronic matter retains vestiges of confinement.  Space-like Wilson
loops still exhibit the area-law behavior associated with confinement.
This is readily seen by considering dimensional reduction, in which
for $T \gg T_c$ the short Euclidean time dimension is collapsed,
leaving three spatial dimensions, one of which is reinterpreted as the
Euclidean time coordinate of a 2+1 dimensional field theory.

The reduction of QCD has these characteristics: 
\bi
\item Quarks acquire a large 3D mass $\sqrt{(\pi T)^2 + m_q^2}$
\item The fourth component of the color vector potential $A_0$ becomes
  a scalar field, and we get a confining gauge-Higgs theory.
\item The 3D and 4D gauge couplings are related through
   $g_3 = g_4 \sqrt{T}$.
\item The spatial Wilson loop gives the 3D potential and 3D string tension.
\ei

\begin{figure}
\begin{center}
\includegraphics[width=0.7\textwidth]{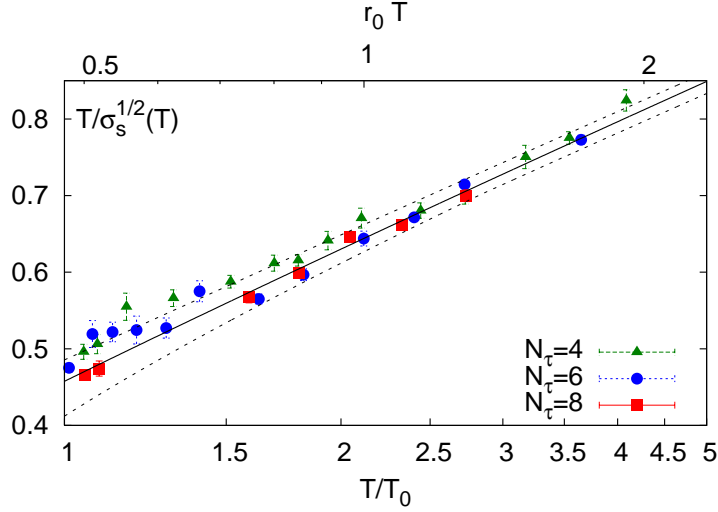} \\
\end{center}
\caption{Temperature divided by the square root of the spatial string
  tension $\sigma_s$ \vs temperature in units of the crossover
  temperature $T_0$ (lower scale) and in $r_0$ units (upper scale) for
  $2+1$ flavors of p4fat3 quarks on lattices with $N_\tau = 4$, 6 and
  8.  The solid curve (with uncertainties indicated by the dashed
  lines) is the prediction of the dimensionally reduced theory
  \protect\cite{Cheng:2008bs}}
\label{fig:spatial_string}
\end{figure}

In a recent calculation Cheng {\it et al} compared the spatial string
tension of the full 4D theory with its predicted behavior in 3D
perturbation theory \cite{Cheng:2008bs}.  The comparison is shown in
Fig.~\ref{fig:spatial_string}. The good agreement with perturbation
theory at temperatures as low as $1.5 T_c$ is unexpected.

\section{Conclusions}

In a reasonably well-matched calculation, new high statistics results
from HotQCD show good agreement between two different staggered
fermion formulations, \ie, p4fat3 and asqtad.  Not surprisingly,
simulations with these inexpensive algorithms are more advanced than
those with other fermion actions, as we have seen from the first
exploratory domain-wall-fermion simulations at $N_\tau = 8$ with a
quite small residual quark mass.  To make progress we need to
understand the importance of cutoff effects and to come closer to the
physical point.  Calculations with other fermion actions can provide
an important check, but those actions must be improved at least to the
same level as the staggered fermion actions before they can play this
role effectively.

We are learning more about the phase structure of zero-baryon-density
QCD as a function of the light quark masses, but these results are
especially sensitive to cutoff effects.  More work is still needed.

There has been recent progress in methodology. The WHOT collaboration
has developed a new method for determining the equation of state, and
Meyer has proposed new methods for determining transport coefficients.

Finally, in measurements of the spatial string tension, we have seen
interesting agreement with predictions of dimensional reduction.

\acknowledgments

I am grateful to my many colleagues for providing figures and
assisting in identifying novel work.  I thank Rajan Gupta, Urs Heller,
and Ludmila Levkova for helpful comments about the manuscript.  This
review is supported by grants from the US Department of Energy and US
National Science Foundation.


\providecommand{\href}[2]{#2}\begingroup\raggedright\endgroup

\end{document}